\documentclass{article}
\usepackage{spconf,amsmath,graphicx}

\usepackage{enumitem}
\setlist{nosep, leftmargin=14pt}

\usepackage{mwe} % to get dummy images

\usepackage{amsmath}
\usepackage{algorithm}
\usepackage{algpseudocode}

\usepackage{graphicx}
\usepackage{multirow}
\usepackage{xcolor}
\usepackage{amssymb}
\usepackage[pagebackref,breaklinks,colorlinks]{hyperref}

\newcommand{\mname}{PMDM-PET}
\title{End-to-End PET Image Reconstruction via a Posterior-Mean Diffusion Model}
\name{Yiran Sun$^{\star, \dagger}$, Osama Mawlawi$^{\dagger, \star}$}
\address{$^{\star}$ Department of Electrical and Computer Engineering, Rice University, Houston, USA\\ $^{\dagger}$ Department of Imaging Physics, University of Texas MD Anderson Cancer Center, Houston, USA}

\begin{document}

\maketitle

\begin{abstract}
Positron Emission Tomography (PET) is a functional imaging modality that enables the visualization of biochemical and physiological processes across various tissues. Recently, deep learning (DL)-based methods have demonstrated significant progress in directly mapping sinograms to PET images. However, regression-based DL models often yield overly smoothed reconstructions lacking of details (i.e., low distortion, low perceptual quality), whereas GAN-based and likelihood-based posterior sampling models tend to introduce undesirable artifacts in predictions (i.e., high distortion, high perceptual quality), limiting their clinical applicability. To achieve a robust perception-distortion tradeoff, we propose \textit{Posterior-Mean Denoising Diffusion Model} (\mname{}), a novel approach that builds upon a recently established mathematical theory to explore the closed-form expression of perception-distortion function in diffusion model space for PET image reconstruction from sinograms. Specifically, \mname{} first obtained posterior-mean PET predictions under minimum mean square error (MSE), then optimally transports the distribution of them to the ground-truth PET images distribution. Experimental results demonstrate that \mname{} not only generates realistic PET images with possible minimum distortion and optimal perceptual quality but also outperforms five recent state-of-the-art (SOTA) DL baselines in both qualitative visual inspection and quantitative pixel-wise metrics PSNR (dB)/SSIM/NRMSE.
\end{abstract}

\section{Introduction}

Positron Emission Tomography (PET) is a functional imaging modality that visualizes biochemical and physiological processes across tissues. During acquisition, high-energy photons from positron annihilation are detected by ring-structured arrays within a PET scanner. Photon counts from multiple angles are compiled into projections and arranged into sinograms, forming the basis for PET image reconstruction~\cite{fahey2002data}. Over the past few decades, various approaches have been developed to reconstruct high-quality PET images from sinograms, broadly categorized into four main directions: 1) Traditional analytical methods, such as filtered back-projection (FBP); 2) Iterative techniques, including ordered subsets expectation maximization (OSEM), maximum-likelihood expectation maximization (MLEM)~\cite{zhu2018ordered}, and penalized log-likelihood (PLL) methods~\cite{wang2012penalized}; 3) Deep learning-based approaches~\cite{haggstrom2019deeppet, whiteley2020directpet, hu2022transem, sun2024r2u, sun2025legopet}; 4) Compressed sensing-based methods~\cite{malczewski2013pet}. Some of these methods have several limitations, for example, traditional analytical approaches are highly sensitive to noise and prone to streak artifacts, iterative methods often suffer from high computational complexity, and compressed sensing techniques rely heavily on the choice of sparsifying transforms and regularization terms.

In this study, we focus on supervised deep learning (DL) methods that directly map sinograms to PET images, as they effectively model complex nonlinear relationships without being constrained by external physical factors. However, regression-based DL approaches trained with mean square error (MSE) often produce overly smooth textures and suboptimal perceptual quality (i.e., low distortion, low perceptual quality)~\cite{blau2018perception}. To improve the perceptual quality in reconstructed images, conditional generative models (cGMs) have emerged as a promising posterior sampling approach, which incorporates novel loss terms to capture underlying posterior distributions and generate realistic images from corresponding measurements. However, cGMs present significant challenges: generative adversarial networks (GANs) often suffer from non-convergence and mode collapse due to their adversarial training process, which may introduce numerous artifacts in predictions~\cite{salimans2016improved}; variational autoencoders (VAEs) struggle with balancing reconstruction fidelity and Kullback–Leibler (KL) divergence, which often results in overly smooth predictions; and denoising diffusion probabilistic models (DDPMs)~\cite{ho2020denoising} involves an iterative denoising process during inference, which sometimes leads to high-frequency details loss in predictions. Furthermore, Freirich et al.~\cite{freirich2021theory} demonstrated that such posterior sampling alone is not an optimal strategy for achieving a desirable perception-distortion tradeoff. Instead, it can be improved by leveraging an optimal MSE estimator without compromising perceptual quality.

To achieve an optimal tradeoff between distortion and perceptual quality, several studies have explored improved reconstruction techniques that minimize distortion while adhering to a specified perceptual constraint~\cite{blau2018perception, freirich2021theory, sun2024r2u, ohayon2024posterior}. Related theoretical frameworks have mathematically demonstrated that this perception-distortion optimization problem can be addressed by optimally transporting posterior mean predictions to the distribution of ground-truth images~\cite{freirich2021theory, sun2024r2u, ohayon2024posterior}. Motivated by this insight, we extend this theory to denoising diffusion probabilistic models (DDPMs) and propose \textit{Posterior-Mean Denoising Diffusion Model} (\mname{}), a simple yet effective framework for generating PET images directly from sinograms while achieving both possible minimum distortion and optimal perceptual quality. Specifically, we first compute posterior mean PET image predictions under MSE distortion supervision. We then employ a conditional diffusion probabilistic model to optimally transport these mean predictions toward the distribution of ground-truth PET images under evidence lower bound (ELBO)-based perception supervision. To evaluate \mname{}, we perform both qualitative and quantitative evaluations in terms of visual inspection and pixel-wise metrics PSNR (dB)/SSIM/NRMSE. Experimental results demonstrate that \mname{} not only generates robust and realistic PET images with a favorable perception-distortion tradeoff but also outperforms five recent state-of-the-art (SOTA) deep learning-based baselines.

\section{Preliminaries}
\label{sec:preliminaries}

\subsection{Perception-Distortion Tradeoff}

Deep learning-based medical image reconstruction approaches are typically evaluated based on two key aspects: (1) the extent to which the reconstructed images \( \hat{\mathbf{x}}_0 \) approximates the reference image \( \mathbf{x}_0 \) on average (distortion quality) and (2) the similarity between the distributions of \( \hat{\mathbf{x}}_0 \) and \( \mathbf{x}_0 \) (perceptual quality). Blau et al.~\cite{blau2018perception} established that distortion and perceptual quality are inherently at odds, where minimizing distortion often comes at the expense of perceptual fidelity, and vice versa. To achieve an optimal tradeoff between perception (P) and distortion (D), we formulate the problem as:
\begin{equation}
    D(P) = \min_{p_{\hat{\mathbf{x}}_0 | \cdot}} \mathbb{E}[\Delta(\mathbf{x}_0, \hat{\mathbf{x}}_0)] \quad \text{s.t.} \quad d(p_{\mathbf{x}_0}, p_{\hat{\mathbf{x}}_0}) \leq P,
\end{equation}
where \( \Delta(\cdot, \cdot) \) represents a distortion measure, and \( d(\cdot, \cdot) \) denotes a divergence between distributions. Specifically, we focus on mean square error (MSE) distortion, defined as \( \Delta(\mathbf{x}_0, \hat{\mathbf{x}}_0) = \|\mathbf{x}_0 - \hat{\mathbf{x}}_0\|^2 \), and minimize it under a perfect perceptual index constraint (\( P=0 \)). This allows us to rewrite Eq.~(1) as:
\begin{equation}
    D(0) = \min_{p_{\hat{\mathbf{x}}_0 |\cdot}} \mathbb{E}[\|\mathbf{x}_0 - \hat{\mathbf{x}}_0\|^2] \quad \text{s.t.} \quad p_{\mathbf{x}_0} = p_{\hat{\mathbf{x}}_0}.
\end{equation}
Recent studies~\cite{freirich2021theory, ohayon2024posterior, sun2024r2u} have shown that the optimal solution to Eq.~(2) can be obtained by first predicting an overly smooth posterior mean \( \mathbf{x}_0^* \) under a possible minimum MSE (MMSE) and then sampling final predictions $\hat{\mathbf{x}}_0$ from the posterior distribution conditioned on \( \mathbf{x}_0^* \). In this work, we extend this theoretical framework to the current state-of-the-art (SOTA) generative modeling paradigm—denoising diffusion probabilistic models (DDPMs)~\cite{ho2020denoising}.

\section{Methods}

\subsection{Overview}
Let \( s \) denote the sinogram images, \( \mathbf{r}_0 \) the corresponding reference PET images, and \( \hat{\mathbf{r}}_0 \) the final predicted PET images, where $s, \mathbf{r}_0, \hat{\mathbf{r}}_0 \in \mathbb{R}^{1 \times H \times W}$. In our setup, the input sinogram images \( s \) and their corresponding PET images \( \mathbf{r}_0 \) are padded to dimensions of \( 1 \times 256 \times 256 \). The general objective of this work is to approximate the stochastic transformation $p({\hat{\mathbf{r}}_0|s})$, i.e., $p(\hat{\mathbf{r}}_0|\mathbf{r}_0^*)p(\mathbf{r}_0^*|s)$, where \(\mathbf{r}_0^*\) represents posterior mean PET predictions obtained from the estimator with possible MMSE distortion \( D^* \). According to~\cite{freirich2021theory}, since \( \hat{\mathbf{r}}_0 \) is independent of \( \mathbf{r}_0 \) given \( s \), the total expected distortion then can be decomposed as:
\begin{equation}
    \mathbb{E}[\|\mathbf{r}_0 - \hat{\mathbf{r}}_0\|^2] = \mathbb{E}[\|\mathbf{r}_0 - \mathbf{r}_0^*\|^2] + \mathbb{E}[\|\mathbf{r}_0^* - \hat{\mathbf{r}}_0\|^2],
\end{equation}
In this work, we optimize the distortion-perception function within the diffusion probabilistic model framework. Then Eq.~(2) and (3) can be reformulated as:
\begin{equation}
\begin{aligned}
D(0) = D^* + \min_{p_{\hat{\mathbf{r}}_0\mathbf{r}_0^*}}
\Bigg\{ 
& \sum_{t=1}^{T} 
    \mathrm{KL}\!\left(
        q(\hat{\mathbf{r}}_t|\hat{\mathbf{r}}_0, \mathbf{r}_0^*) 
        \parallel 
        p_\theta(\hat{\mathbf{r}}_t|\hat{\mathbf{r}}_{t+1}, \mathbf{r}_0^*)
    \right)
\\
& :\;
p_{\hat{\mathbf{r}}_0\mathbf{r}_0^*} \in 
    \mathcal{M}(p_{\mathbf{r}_0}, p_{\mathbf{r}_0^*})
\Bigg\}.
\end{aligned}
\end{equation}
where \( \mathcal{M}(p_{\mathbf{r}_0}, p_{\mathbf{r}_0^*}) \) is the set of all probability distributions on \( \mathbb{R}^{n_{\mathbf{r}_0}} \times \mathbb{R}^{n_{\mathbf{r}_0}} \) with marginals \( p_{\mathbf{r}_0} \) and \( p_{\mathbf{r}_0^*} \). We note that this formulation actually aims to minimize the sum of Kullback–Leibler (KL) divergences over \( T \) timesteps between \( p_{\mathbf{r}_0} \) and \( p_{\mathbf{r}_0^*} \), given an estimator with possible MMSE distortion \( D^* \). Therefore, based on this nature, our proposed method \name{} consists of two stages: 1) train a regression-based DL model to predict the posterior mean PET images \( \mathbf{r}_0^* \) by minimizing the MSE between reconstructed and ground-truth images (see Sec.~2.1), 2) train a conditional diffusion model to learn the optimal transport path between the distributions of posterior-mean predictions and ground-truth images.  
The following sections provide a detailed description of \mname{}.

\subsection{MSE Supervised Posterior-Mean Estimator}

For consistency in comparison, we adopt DeepPET~\cite{haggstrom2019deeppet}, a convolutional encoder-decoder network, as our MSE-based estimator model \( f_{\phi} \) for $p(\mathbf{r}_0^*|s)$. The model takes sinograms $s$ as input and outputs overly smoothed posterior mean PET images \( \mathbf{r}_0^* \). The encoder consists of convolutional blocks with a stride of 2, doubling the number of feature maps at each stage. Filter sizes decrease from \( 7 \times 7 \) (first two layers) to \( 5 \times 5 \) (next two layers) and \( 3 \times 3 \) for the remaining layers, producing a final output of 512 feature maps at a resolution of \( 32 \times 32 \). The decoder progressively upsamples the representation using interpolation, followed by \( 3 \times 3 \) convolutions that halve the number of feature maps, batch normalization (BN), and ReLU activations. The full network comprises 22 convolutional layers. Alternative neural network designs may achieve comparable performance. The estimator \( f_{\phi} \) is trained end-to-end under MSE supervision on sinogram-PET pairs:  
\begin{equation}
    \mathcal{L}_{DeepPET} := \mathbb{E} \left[ \|\mathbf{r}_0 - f_{\phi}(s)\|^2 \right].
\end{equation}
The possible MMSE distortion \( D^* \) is then attained by predicting posterior mean PET images \( \mathbf{r}_0^* \) from the trained estimator model \( f_{\phi^*}(s) \).

\subsection{Posterior-Mean Denoising Diffusion Model} 

\mname{} approximates the posterior-mean conditioned distribution $p(\hat{\mathbf{r}}_0|\mathbf{r}_0^*)$ through a fixed forward process and a learning-based reverse process. The deterministic forward process begins with clean PET samples drawn from the input data distribution \(\mathbf{r}_0 \sim q(\mathbf{r}_0) \) and \(\mathbf{r}_t \) is a linear combination of noise \( \epsilon \sim \mathcal{N}(0, \mathbf{I}) \) and \( \mathbf{r}_0 \), i.e. $\mathbf{r}_t = \sqrt{\overline{\alpha}_t}\mathbf{r}_0 + \sqrt{1-\overline{\alpha}_t}\epsilon$. The goal of the learning-based reverse process is to generate clean PET images \( \hat{\mathbf{r}}_0 \) from the noisy sample \(\mathbf{r}_T\sim\mathcal{N}(0, \mathbf{I})\) given posterior mean PET images \(\mathbf{r}_0^*\), i.e. approximate the optimal transport $p(\hat{\mathbf{r}}_0|\mathbf{r}_0^*)$. This process is modeled using a joint Markov chain $p_{\theta}(\mathbf{r}_{0:T}) := p(\mathbf{r}_T)\prod_{t=1}^T p_{\theta}(\mathbf{r}_{t-1} | \mathbf{r}_t, \mathbf{r}_0^*)$, and the transition probabilities are parameterized as $\mathcal{N}(\mathbf{r}_{t-1}; \mu_{\theta}(\mathbf{r}_t, t, \mathbf{r}_0^*), \textstyle\sum\nolimits_{\theta}(\mathbf{r}_t, t, \mathbf{r}_0^*))$, where \(\theta \) represents the learnable parameters of the neural network. The mean function is further reparameterized as $\mu_{\theta}(\cdot, \cdot) = \frac{1}{\sqrt{\alpha_t}}\left(\mathbf{r}_t - \frac{1-\alpha_t}{\sqrt{1-\bar{\alpha}_t}}\epsilon_{\theta}(\mathbf{r}_t, t, \mathbf{r}_0^*)\right)$, where \( \epsilon_{\theta}(\cdot, \cdot) \) predicts the noise added at each time step.  

Align with previous work on diffusion models~\cite{ho2020denoising}, we utilize a convolutional U-Net for denoising at each step of the reverse process. The posterior mean estimation \(\mathbf{r}_0^* \) is incorporated into the denoising process as an additional condition. During training, at each iteration, a batch of random sinogram-PET pairs is sampled, and \mname{} is optimized by minimizing the denoising loss (where $\mathbf{r}_0^*=f_{\phi^*}(s)$):
\begin{equation}
    \mathcal{L}_{\name{}} := \mathbb{E}_{t\sim [1,T],\mathbf{r}_0,\epsilon}\left[||\epsilon_t - \epsilon_{\theta}(\mathbf{r}_t,t,\mathbf{r}_0^*)||^2\right].
\end{equation}
During inference, starting from Gaussian noise $\mathbf{r}_T$, we iteratively compute $\mathbf{r}_{t-1} = \tfrac{1}{\sqrt{\alpha_t}}\!\left(\mathbf{r}_t - \tfrac{\beta_t}{\sqrt{1-\bar{\alpha}_t}}\epsilon_\theta(\mathbf{r}_t, t, \mathbf{r}_0^*)\right) + \sigma_t\mathbf{z}$ over $T$ steps to generate clean predicted images $\hat{\mathbf{r}}_0$, where $\mathbf{z}\!\sim\!\mathcal{N}(0,\mathbf{I})$.

\section{Experiments}
\begin{figure*}[t!]
    \centering
    \includegraphics[width=0.765\textwidth]{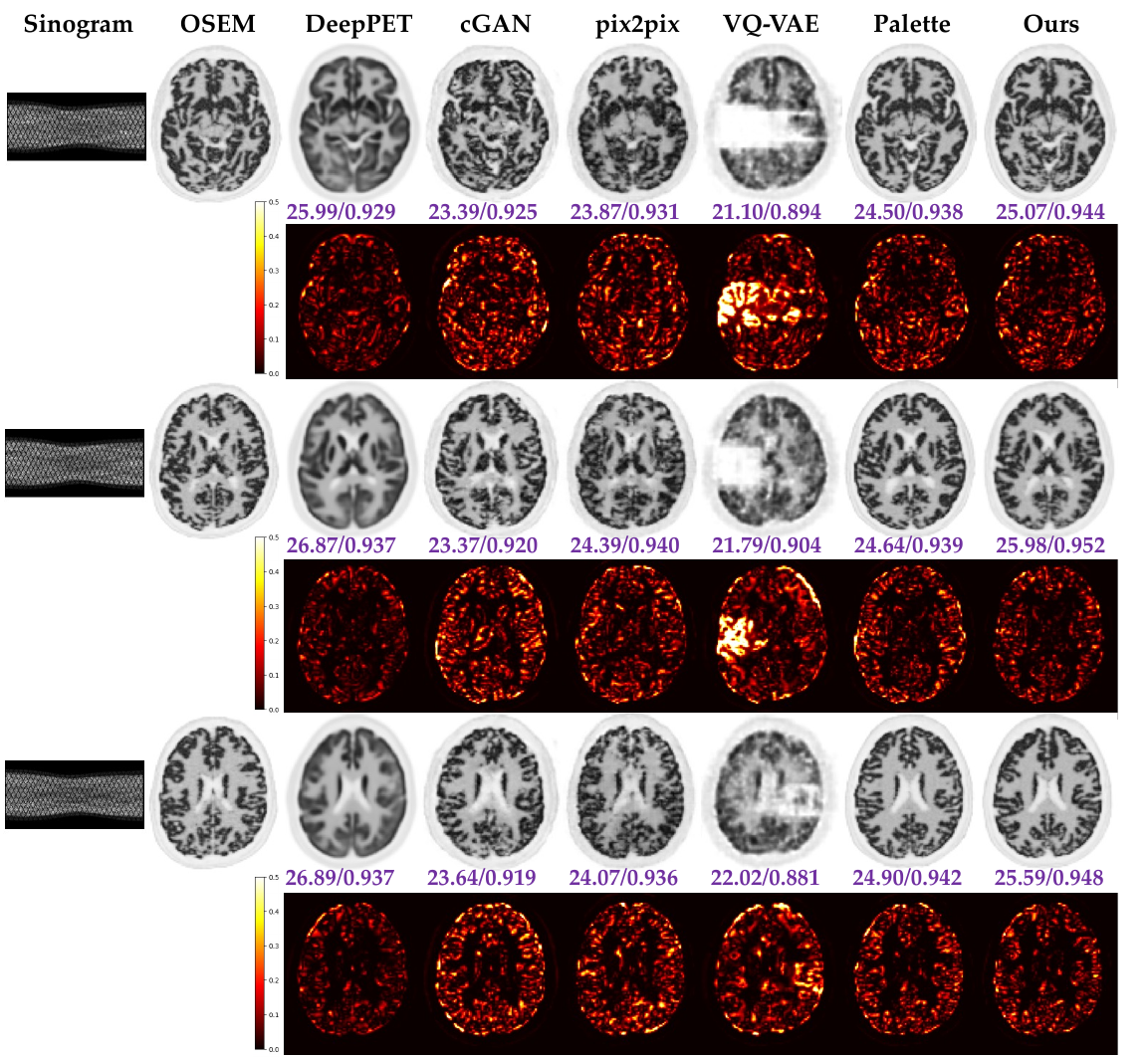}
    \caption{\textbf{Qualitative comparison of \mname{} with Five Baselines on Three Example Slices.} The first column shows the input sinogram images, and the second column shows the reference images reconstructed using OSEM algorithm. The third to seventh columns correspond to the five baselines (labeled above each image), and the final column shows the reconstructed PET images using our proposed \mname{} method. PSNR/SSIM values are reported below each slice, and squared error maps between each method and the reference image are also displayed (second, fourth and sixth rows).}
    \label{fig:visual}
\end{figure*}

\subsection{Datasets and Preprocessing}
We follow the data processing steps and use the code from~\cite{mehranian2020model} for sinogram-PET pair data preparation. We simulated 2D $^{18}$F-FDG PET images from 20 BrainWeb 3D brain phantoms~\cite{collins1998design}, acquired on a Siemens Biograph mMR scanner with a spatial resolution of $2.086 \times 2.086 \times 2.031$ mm$^3$ and a matrix size of $344 \times 344 \times 127$. To augment the dataset, each phantom was axially rotated by five random angles in the range $[0, 15]$ degrees, yielding $20*5$ 3D brain phantoms. For each phantom, we extracted 55 non-contiguous axial slices and generated their corresponding sinograms using $1 \times 10^{10}$ counts and point spread function (PSF) modeled with 2.5 mm full-width-at-half-maximum (FWHM) Gaussian kernels. High-count reference PET images were reconstructed using OSEM algorithm with 10 iterations and 14 subsets. The dataset was divided into $17*5$ brain phantoms (4675 slices) for training, $1*5$ brain phantoms (275 slices) for validation, and $2*5$ brain phantoms (550 slices) for testing.

\subsection{Implementation Details}
All experiments were implemented in PyTorch and run on NVIDIA Quadro RTX 8000 GPUs. The MSE estimator (Sec.~3.2) was trained with Adam optimizer (learning rate $1\times10^{-4}$, weight decay $1\times10^{-5}$). \mname{} was then trained with a learning rate of $3\times10^{-5}$. Both models were trained for 500 epochs with a batch size of 4. Reconstruction quality was evaluated using PSNR and NRMSE for distortion, and SSIM plus visual inspection for perceptual quality.

\subsection{Reconstruction Results}

\begin{table}[t!]
\centering
\setlength{\tabcolsep}{3pt}
\renewcommand{\arraystretch}{1.0}
\footnotesize 
\caption{\textbf{Quantitative Evaluation using PSNR (dB)/SSIM/NRMSE.} 
Red and blue indicate the best and second-best results.}
\label{tab:comparison}
\vspace{2pt}

\begin{tabular}{l|l|c|c|c}
\hline
\multirow{2}{*}{\textbf{Type}} & \multirow{2}{*}{\textbf{Model}} 
& \multicolumn{3}{c}{\textbf{Metrics}} \\ \cline{3-5}
 &  & PSNR$\uparrow$ & SSIM$\uparrow$ & NRMSE$\downarrow$ \\ 
\hline
\multirow{1}{*}{Regression-based} 
& DeepPET~\cite{haggstrom2019deeppet} 
& \textcolor{red}{\textbf{27.10 $\pm$1.01}} 
& 0.938 $\pm$ 0.007 
& \textcolor{red}{\textbf{0.044}} \\
\hline
\multirow{2}{*}{GAN-based} 
& cGAN~\cite{mirza2014conditional} 
& 22.20 $\pm$ 3.12 
& 0.909 $\pm$ 0.035 
& 0.084 \\
& pix2pix~\cite{isola2017image} 
& 22.43 $\pm$ 3.25 
& 0.915 $\pm$ 0.036 
& 0.082 \\
\hline
\multirow{3}{*}{Likelihood-based} 
& VQ-VAE~\cite{van2017neural} 
& 20.86 $\pm$ 2.16 
& 0.873 $\pm$ 0.023 
& 0.093 \\
& Palette~\cite{saharia2022palette} 
& 25.40 $\pm$ 1.00 
& \textcolor{blue}{\textbf{0.945 $\pm$ 0.007}} 
& 0.054 \\
& \textbf{\mname{}} 
& \textcolor{blue}{\textbf{26.01 $\pm$ 1.03}} 
& \textcolor{red}{\textbf{0.950 $\pm$ 0.006}} 
& \textcolor{blue}{\textbf{0.050}} \\
\hline
\end{tabular}
\vspace{-6pt}
\end{table}

We compared the performance of \mname{} against five baseline algorithms: DeepPET~\cite{haggstrom2019deeppet}, cGAN~\cite{mirza2014conditional}, pix2pix~\cite{isola2017image}, VQ-VAE~\cite{van2017neural}, and Palette~\cite{saharia2022palette}. It is worth noting that DeepPET can be regarded as a model trained without a perceptual constraint and Palette is trained without an MSE estimator, which can also be seen as an ablation study when compared to our \mname{}. Table.~\ref{tab:comparison} summarizes the average pixel-wise PSNR (dB)/SSIM/NRMSE values across all reconstructed slices, along with the number of trainable parameters for each method. \mname{} achieves a 0.61 dB improvement in PSNR over Palette, demonstrating the effectiveness of incorporating an MMSE estimator to reduce distortion while preserving high perceptual quality. While DeepPET achieves comparable PSNR values, its reconstructed PET images appear significantly blurrier and less perceptually faithful, as illustrated in Fig.~\ref{fig:visual}. This aligns with the tradeoff theory discussed in Sec.~2.1, where generative models leveraging stochastic posterior sampling sacrifice pixel-wise distortion (resulting in lower PSNR) while maintaining fidelity to the underlying distribution~\cite{blau2018perception}. Additionally, \mname{} outperforms cGAN, pix2pix, and VQ-VAE by approximately 3.81 dB, 3.58 dB, and 5.15 dB in PSNR, respectively. Fig.~\ref{fig:visual} further visualizes reconstruction results, comparing all baseline methods side by side. \mname{} produces the most robust reconstructions, excelling in both distortion and perceptual quality. In contrast, cGAN, pix2pix, VQ-VAE, and Palette introduce noticeable artifacts, while DeepPET generates overly smooth reconstructions with unrealistic fine details.

\section{Conclusion}

We propose \textit{Posterior-Mean Denoising Diffusion Model} (\mname{}) to generate PET images directly from sinograms while achieving a favorable balance between distortion and perceptual quality. We evaluate \mname{} on simulated human brain data~\cite{collins1998design} using both qualitative and quantitative metrics. Experimental results demonstrate that \mname{} produces highly realistic PET images with an optimal distortion-perception tradeoff and outperforms five recent SOTA DL baselines. Future work will focus on clinical evaluations.

\section{COMPLIANCE WITH ETHICAL STANDARDS}
This is a numerical simulation study for which no ethical approval was required.

\section{Acknowledgments}
The authors declare no conflicts of interest.

\bibliographystyle{IEEEbib}
\bibliography{strings,refs}

\end{document}